\documentclass[10pt,twocolumn,reprint,superscriptaddress,amsmath,amssymb,aps]{revtex4-2}
\usepackage{times}
\usepackage{graphicx}% Include figure files
\usepackage{dcolumn}% Align table columns on decimal point
\usepackage{bm}% bold math
\usepackage{amsmath}
\usepackage{xr-hyper}
\usepackage[colorlinks=true, allcolors=blue]{hyperref}
\usepackage{ulem} % strike out 
\usepackage{enumitem}
\usepackage[version=3]{mhchem}

\begin{document}

\title{Influence of graphene on the electronic and magnetic properties of an iron(III) porphyrin chloride complex}

\author{Young-Joon Song}
\email{ysong@itp.uni-frankfurt.de}
\affiliation{Institut f\"ur Theoretische Physik, Goethe-Universit\"at Frankfurt, Max-von-Laue-Str. 1, 60438 Frankfurt am Main, Germany}

\author{Charlotte Gallenkamp}
\affiliation{TU Darmstadt, Department of Chemistry, Quantum Chemistry, Peter-Gr\"unberg-Str. 4, 64287 Darmstadt, Germany}

\author{Gen\'{i}s Lleopart} 
\affiliation{Departament de Ci\'encia de Materials i Qu\'imica F\'isica and Institut de Qu\'imica Te\'orica i Computacional (IQTC), Universitat de Barcelona, c/ Mart\'i i Franqu\'es 1-11, 08028 Barcelona, Spain}

\author{Vera Krewald}
\email{vera.krewald@tu-darmstadt.de}
\affiliation{TU Darmstadt, Department of Chemistry, Quantum Chemistry, Peter-Gr\"unberg-Str. 4, 64287 Darmstadt, Germany}

\author{Roser Valent\'{i}}
\email{valenti@itp.uni-frankfurt.de}
\affiliation{Institut f\"ur Theoretische Physik, Goethe-Universit\"at Frankfurt, Max-von-Laue-Str. 1, 60438 Frankfurt am Main, Germany}

\date{\today}

\begin{abstract}
Although iron-based single atom catalysts are regarded as a promising alternative to precious metal catalysts, their precise electronic structures during catalysis still pose challenges for computational descriptions. A particularly urgent question is the influence of the environment on the electronic structure, and how to describe this properly with computational methods. Here, we study an iron porphyrin chloride complex adsorbed on a graphene sheet using density functional theory calculations to probe how much the electronic structure is influenced by the presence of a graphene layer. 
Our results indicate that weak interactions due to van der Waals forces dominate between the porphyrin complex and graphene, and only a small amount of charge is transferred between the two entities. Furthermore, the interplay of the ligand field environment, strong $p$ $-$ $d$ hybridization, and correlation effects within the complex are strongly involved in determining the spin state of the iron ion. By bridging molecular chemistry and solid state physics, this study provides first steps towards a joint analysis of the properties of iron-based catalysts from first principles. 
\end{abstract}
\maketitle

\section{Introduction}
Iron as a versatile element for catalysis has received increasing attention over the past years. When bound in a macrocycle such as phthalocyanine or porphyrin, it can serve as a model for bioinorganic or man-made active sites. \cite{zhao_2019,anxolabehere-mallart_2019,amanullah_2019,huang_2018} A field that has grown particularly rapidly is single-atom catalysis, where such types of active site are embedded in an extended graphene sheet or nanotube or carbon-based environments that form during a pyrolysis step. For instance, in FeNC catalysts, the active sites are commonly discussed as a single iron ion coordinated by four nitrogen donors (\ce{FeN4}). \cite{zitolo_2015,li_2016,ni_2021,li_2020,ni_2022,kumar_2023,mineva_2019} Additionally, the heterogenisation of iron complexes for applications in electrocatalysis, for instance by physisorption, covalent linkage, or incorporation in a conductive polymer, has become an important modification of the environment of these complexes with implications for catalysis.\cite{heppe_2023}

\begin{figure*}[t]
\centering\includegraphics[width=1.7\columnwidth]{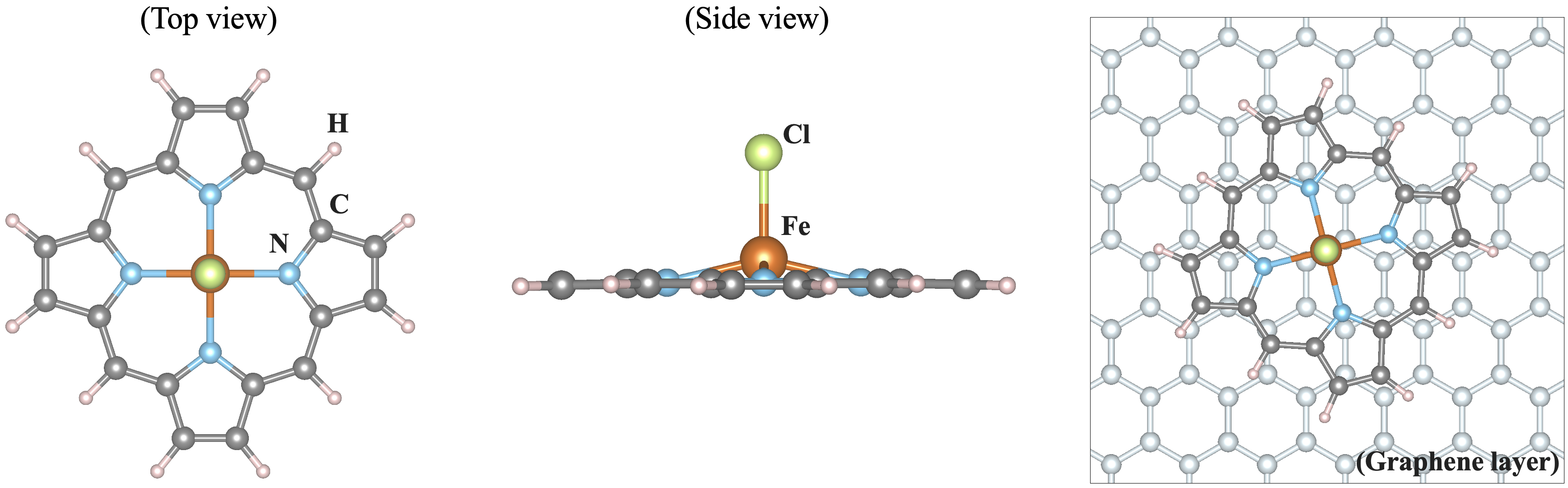}
\caption{
(Left, Middle) Crystal structure of the truncated iron porphyrin chloride complex \ce{[Fe^{III}(P)(Cl)]}.
(Right) The resulting optimized structure of "bridge02" which is the most energetically stable case among the four considered for this study (see main text). The center of the \ce{[Fe^{III}(P)(Cl)]} complex is situated above a C-C bond of the graphene layer.}
\label{Fe-P_str}
\end{figure*}

For computational models of single-atom catalysts or iron complexes incorporated into electrode materials, the question arises how the environment should be described. An important component of the environment in electrocatalysts is the often carbon-based material of the electrode, raising the question of the number of graphene layers that should be considered explicitly. Herein, we address this aspect by comparing the electronic structure of an isolated, neutral iron porphyrin complex, \ce{[Fe^{III}(P)(Cl)]} (P: porphyrin ligand; Cl: chloride ligand), with that of the same complex adsorbed on an extended graphene sheet, see Figure \ref{Fe-P_str}. Since the electronic structure of the isolated complex is well-known, the effect of the graphene sheet can be evaluated well. We find that the influence of a graphene layer parallel to the catalyst plane on the electronic structure of \ce{[Fe^{III}(P)(Cl)]} is negligible, implying that future catalytic models can concentrate on single-layer models. 

This evaluation is complemented by a comparison of the electronic structure descriptions with molecular and periodic approaches. Particular attention is paid to different spin states and their relative stabilities in these two types of description. We highlight that, when aiming to describe Fe-based catalyst models with computational methods, theoreticians can use periodic methods with plane-wave basis sets and extended structural models or molecular methods with Gaussian basis sets and finite structural models. \cite{ni_2021,ni_2022,mineva_2019,kattel_2014} These two approaches differ not only in the model size, but also in the electronic structure treatments and underlying assumptions. This inhibits the comparability of results obtained with different approaches. We propose here how these descriptions may be compared since both treatments can be used in complementary ways to model the same type of catalyst or material.

\section{Methods}

\textbf{Methodology}. The structure of iron(III) in an equatorial octaethylporphyrin (OEP) ligand sphere and an axially bound chloride ligand serves as a reference, with its coordinates as given in the Supplementary Information. The truncated \ce{[Fe^{III}(P)(Cl)]} complex is obtained by replacing the eight ethyl substituents on the porphyrin ligand by hydrogen atoms and relaxing the resulting structure. The molecular calculations were performed within density functional theory (DFT) using the
Tao–Perdew–Staroverov–Scuseria (TPSS) exchange-correlation functional \cite{tao_climbing_2003} in ORCA 5.0.4 \cite{orca_2012,orca_v5} with the Gaussian basis sets def2-TZVP on Fe, N and Cl and def2-SVP on C and H (labeled ef2-TZVP:def2-SVP). \cite{weigend_balanced_2005} Convergence criteria for the geometry relaxation and self-consistent-field
(SCF) convergence were set to `Tight' in ORCA nomenclature. The dispersion correction D3BJ\cite{grimme_consistent_2010,grimme_effect_2011} and implicit solvation model SMD with water\cite{marenich_universal_2009} as a solvent were used.
For the electronic structures, single point calculations were carried out with the OPBE exchange-correlation functional  and the CP(PPP) \cite{neese_prediction_2002} basis set on Fe, while def2-TZVP \cite{weigend_balanced_2005} was employed for all other atoms. No dispersion correction was used for the single point calculations. These choices were made based on the literature on spin state prediction for iron complexes\cite{vlahovic_density_2019,phung_toward_2018,radon_benchmarking_2019} and in-house calibration studies for \ce{FeN4} environments in particular.

The geometry and electronic structure of the truncated \ce{[Fe^{III}(P)(Cl)]} complex adsorbed on a graphene sheet was investigated with DFT using the projector augmented wave method\cite{paw} implemented in the Vienna \textit{ab initio} simulation package (VASP)\cite{vasp} with periodic boundary conditions. The generalized gradient approximation (GGA) was employed for the exchange-correlation functional. \cite{gga} The size of the basis set was determined by the energy cutoff of 500 eV and the van der Waals correction (DFT-D2)\cite{vdW-d2} was taken into account in the calculations. To treat localized Fe 3$d$ orbitals properly and account for correlation effects, an onsite $U$ and Hund's coupling $J_H$ (GGA+$U$) were introduced.\cite{Liechtenstein_1995} As discussed below in the results, the preferred spin state of iron depends on the choice of $U$, and $U$ = 4 eV and $J_H$ = 1 eV are employed in the calculations to obtain the experimentally observed high spin configuration. 
We also utilized the meta-GGA functional r2SCAN without including $U$ in the calculations. \cite{scan,r2scan}
For the structure optimisation, the coordinates of the graphene sheet were taken from a single layer of graphite with a C-C bond length of 1.42 \AA. The \ce{[Fe^{III}(P)(Cl)]} complex was placed in different positions on the graphene sheet (see below). With the GGA + $U$ description, all atoms of the \ce{[Fe^{III}(P)(Cl)]} complex were fully relaxed until the net force was smaller than 10$^{-3}$ eV/\AA, while the graphene sheet was kept fixed. 
Additionally, the nonlocal van der Waals correction was tested during structure optimisation using the r2SCAN+rVV10 nonlocal vdW-DF functional implemented in VASP. \cite{vdW_functional,r2SCAN+rVV10}
The self-consistent field procedure used convergence criteria of 10$^{-6}$ eV and the 6$\times$6$\times$1 $k$ mesh.
To simulate the \ce{[Fe^{III}(P)(Cl)]} complex without graphene in VASP, a large unit cell is utilized, where two iron atoms between adjacent unit cells are separated by a distance of 20 \AA, using only one $k$ point.
Some figures were visualized using VESTA \cite{vesta} and PyProcar.\cite{pyprocar1,pyprocar2}

\section{Results}
\textbf{Analysis of the isolated molecular complex.} 
The geometry of \ce{[Fe^{III}(OEP)(Cl)]} optimised with TPSS/def2-TZVP:def2-SVP in the expected high spin electronic configuration\cite{kintner_1991} has structural parameters consistent with experiment (calc: d(Fe,Cl) = 2.264 \AA, d(Fe,N\textsubscript{av}) = 2.088 \AA; exp: d(Fe,Cl) = 2.225(4) \AA, d(Fe,N\textsubscript{av}) = 2.065(2) \AA \cite{kohnhorst_2014}) and a small displacement of the iron(III) ion from the plane spanned by the four nitrogen atoms (calc: d(Fe,plane) = 0.466 \AA; exp: d(Fe,plane) = 0.494(4) \AA). Energetically, the high spin structure is found to be degenerate with the structure optimised as the intermediate spin case ($<1.0$ kcal/mol using OPBE/CP(PPP):def2-TZVP), while the low spin structure is clearly disfavoured (15.8 kcal/mol). 

To facilitate faster calculations in the graphene adsorption studies, the eight ethyl substitutents were truncated to hydrogen atoms, resulting in the unsubstituted porphyrin complex \ce{[Fe^{III}(P)(Cl)]}. The truncation has negligible effects on the structure and spin state splittings. The bond distance changes are predicted to be minimal (calc: d(Fe,Cl) = 2.248 \AA, d(Fe,N\textsubscript{av}) = 2.090 \AA). Similarly, the energetic ordering of spin states is unchanged, i.e. degenerate high and intermediate spin states ($<1.0$ kcal/mol) and disfavoured low spin state (+15.2 kcal/mol). Relaxing this structure in VASP leads to very small structural changes. The Fe-Cl distance is predicted at 2.208 \AA, and the average Fe-N distance is found at 2.085 \AA. 
To investigate the bond distance in hypothetical low and intermediate spin states of iron, the \ce{[Fe^{III}(P)(Cl)]} structure is additionally relaxed in VASP by fixing the total spin moment to be 1 $\mu_{B}$ and 3 $\mu_{B}$ representing a low and intermediate spin configuration respectively. The resulting average bond distance between Fe and N is measured as 1.994 \AA~(2.016 \AA) in the low (intermediate) spin state, i.e. as expected shorter than in the high spin state.

\begin{figure}[t]
\vskip 2mm
\includegraphics[width=0.9\columnwidth]{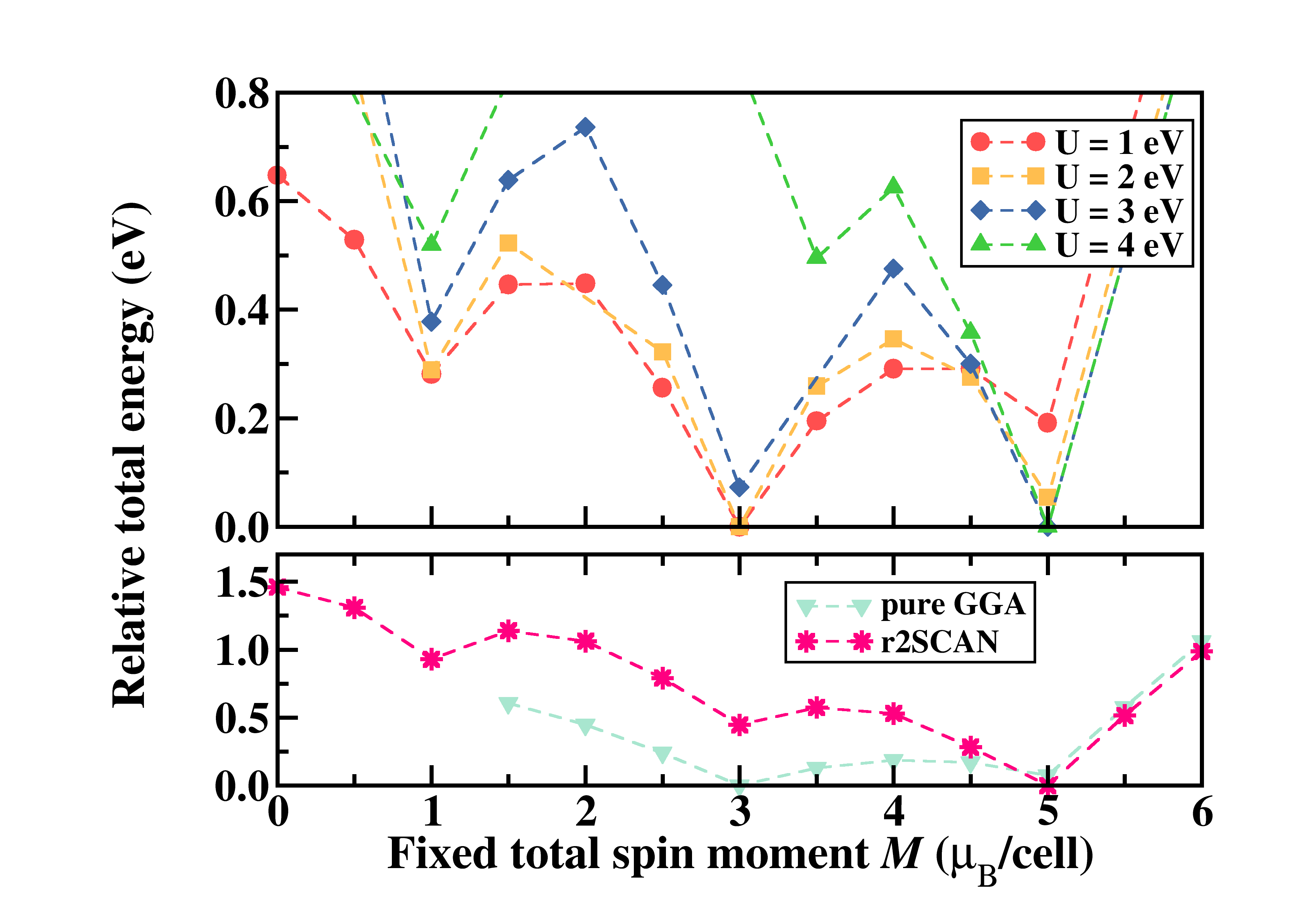}
\caption{
Total energy in \ce{[Fe^{III}(P)(Cl)]} as a function of fixed spin moments using GGA+$U$ and meta-GGA (r2SCAN) in VASP. Each minimum energy is chosen for the energy zero. In the case of GGA+$U$, a high spin (S=5/2) configuration can be obtained at $U$ larger than 3 eV.
}
\label{Fe-P_fsm}
\end{figure}

Similar to the spin state energetics found for \ce{[Fe^{III}(P)(Cl)]} in ORCA, the VASP results in \ce{[Fe^{III}(P)(Cl)]} also reveal that the low spin structure is energetically unfavored with all exchange-correlation functionals used here, as shown in Figure \ref{Fe-P_fsm} that illustrates relative total energy as a function of the total spin moment obtained via fixed spin moment calculations. 
It is evident that $U$ $\ge$ 3 eV in GGA + $U$ leads to a high spin iron species, whereas $U$ $<$ 3 eV favors the intermediate spin configuration of iron. Furthermore, the parameter-free meta-GGA r2SCAN stabilises the complex in a high spin configuration of iron.
Specifically, the energy difference between the intermediate and high spin states of iron is obtained to be about 854 meV (19.7 kcal/mol) in GGA+$U$(4 eV) and 446 meV (10.3 kcal/mol) in r2SCAN. These energy differences are larger than those predicted with the molecular approach (vide supra). 

Numerous studies have focused on iron porphyrin complexes and similar systems, employing diverse theoretical approaches to unveil the ground state spin configuration of iron. \cite{berryman_balancing_2015,nachtigallova_isolated_2018,antalik_ground_2020} Our results are in agreement with these studies, including the predicted high spin ground state and relaxed Fe-N distance of 2.085 Å for \ce{[Fe^{III}(P)(Cl)]} which closely aligns with the range associated with high-spin configurations. 
Our focus therefore turns to the interaction of the complex with a carbon-based environment, represented here by a graphene sheet as this type of environment is expected in \ce{FeN4} catalyst materials.

\begin{figure}
\vskip 2mm
\includegraphics[width=1\columnwidth]{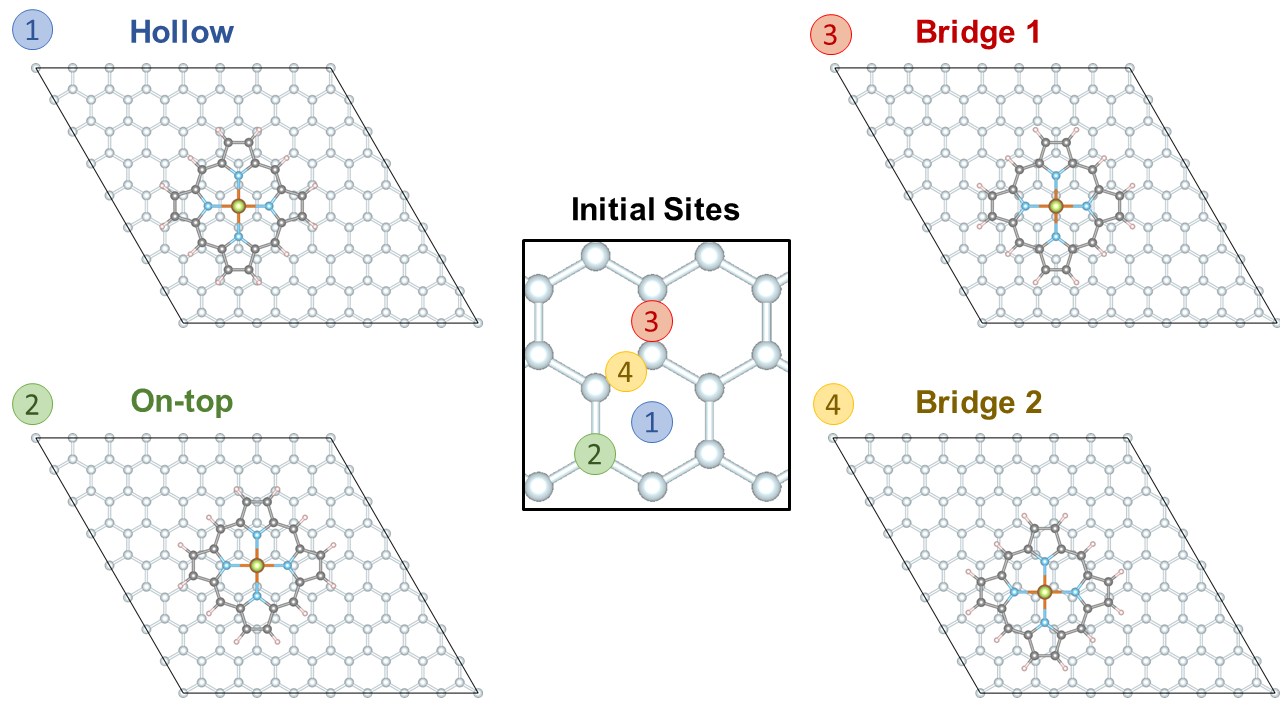}
\caption{
Structures of the \ce{[Fe^{III}(P)(Cl)]} complex adsorbed on a graphene layer considering four different initial sites of graphene for structural relaxation.
}
\label{Fe-P_rel}
\end{figure}

\textbf{Interaction with a graphene surface}.
To evaluate the interaction of [Fe(P)(Cl)] with a graphene sheet, as shown in Figure \ref{Fe-P_str}, the molecule was deposited on graphene sheets of sizes 8$\times$8 and 6$\times$6. 
We find small differences in the results for the 8$\times$8 and 6$\times$6 graphene cases. While we here focus on the 8$\times$8 graphene sheet representation since computational cost was not a concern, future studies may use the smaller 6$\times$6 graphene sheet size. 
Four possible interaction sites were considered as starting points: (i) ``hollow'', where the centers of the iron complex and a central benzene ring align, (ii) ``on-top'', where the center of the iron complex is placed directly above a carbon atom, (iii) ``bridge 1'' and (iv) ``bridge 2'', where the center of the iron complex is placed above a C-C bond, see Figure \ref{Fe-P_rel}. 
Relaxation of the structures shows that the porphyrin complex in the site ``on-top'' moves and becomes the same as ``bridge 1''. For ``bridge 2'' the porphyrin complex does not move, but rotates by about 15 degrees. Contrary to the above cases, the positioning of the porphyrin complex remains unchanged in the case ``hollow'', see the Supplementary Information. 

Our results show that the ``bridge 2'' case is the most stable structure, but the energetic difference between the three sites is not substantial. The final structures ``bridge 1'' and ``hollow'' are found ca. 24 meV and 47 meV above the minimum structure ``bridge 2'', i.e. less than 1.1 kcal/mol. 
These results demonstrate that the interaction between \ce{[Fe^{III}(P)(Cl)]} is very small and isotropic with respect to the graphene plane. 
This agrees well with an experimental study of a single iron(II) phthalocyanine complex adsorbed on graphene, indicating that its position can be easily manipulated.\cite{de_la_torre_non-covalent_2018}

\begin{figure}
\vskip 2mm
\includegraphics[width=1.0\columnwidth]{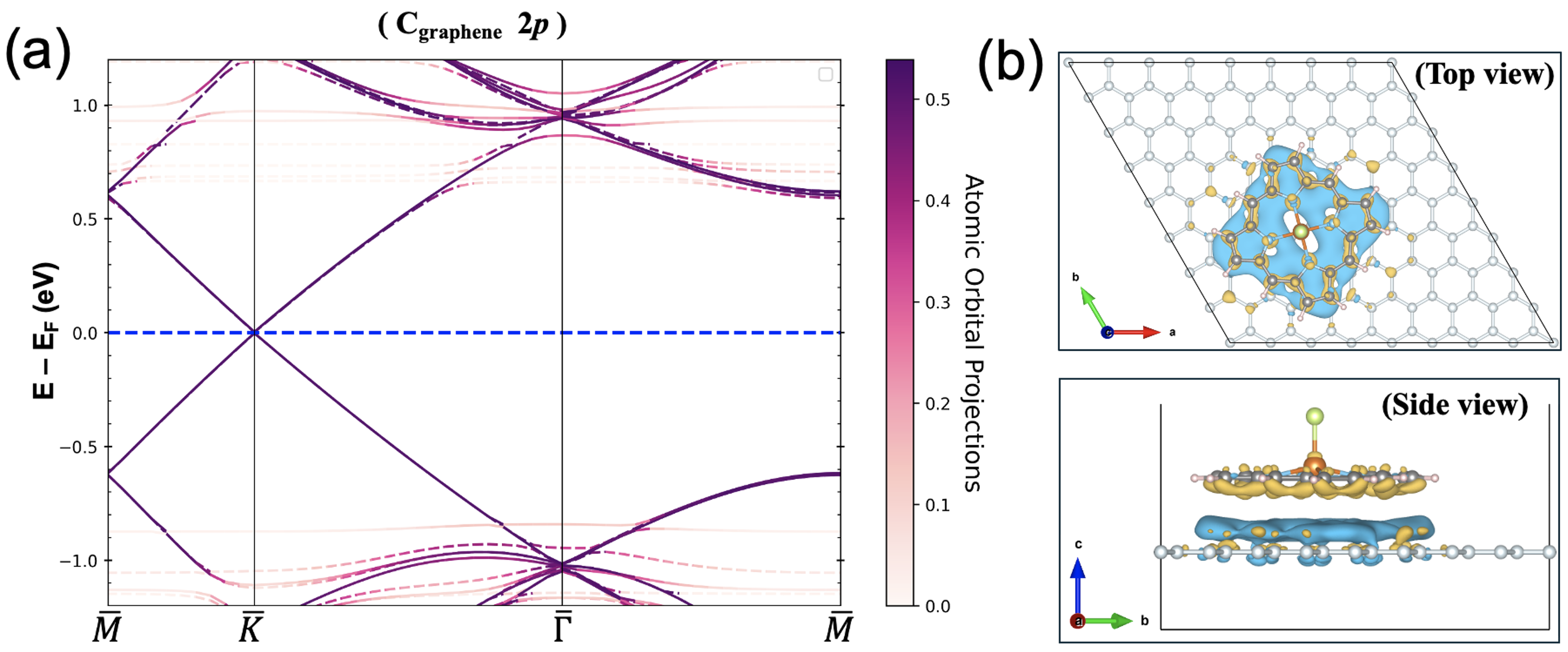}
\caption{
(a) GGA+$U$(4 eV) spin-polarized C$_{graphene}$ 2$p$ orbitals projected band structures in the \ce{[Fe^{III}(P)(Cl)]} complex adsorbed on a graphene sheet (8$\times$8) along the high symmetry points where solid (dashed) lines represent spin-up (spin-down) bands.  The Fermi energy is set to zero. 
(b) The corresponding charge difference plot with an isovalue of 0.0013 $e$/\AA. Charge depletion and accumulation regions are visualized in yellow and light blue, respectively.}
\label{Fe-P_bands_cd}
\end{figure}

The minimal interaction is also clear from a Bader charge analysis, where the charge transfer within GGA + $U$ amounts to 3.9 $\times$ 10$^{-3}$ $e$/molecule. This can be seen in Fig. \ref{Fe-P_bands_cd}(a) where the GGA + $U$  C$_{graphene}$ 2$p$ orbital-resolved band structure of the system is depicted along the high symmetry points.
Orbital-resolved band structures for the remaining atoms are shown in the Supplementary Information.
The Dirac point at the K point coming from the graphene emerges at the Fermi energy without any shifts resulting from the small interaction. Figure \ref{Fe-P_bands_cd}(b) visualizes the corresponding charge difference plot, defined as $\rho_{AB}$ $-$ $\rho_A$ $-$ $\rho_B$, showing where regions of charge accumulation (light blue) and depletion (yellow) are localized.
Due to the small interaction, spin density plots, defined as $\rho_\uparrow$ $-$ $\rho_\downarrow$, within GGA + $U$ are very much the same with and without graphene in the system, see Figure S\ref{supp-Fe-P_sd} in the Supplementary Information. 
At the GGA level, the adsorption energy, defined as $E_{ads}$ = $E_{AB}$ $-$ $E_A$ $-$ $E_B$ where $A$ and $B$ represent \ce{[Fe^{III}(P)(Cl)]} and graphene respectively, is calculated as $-$1.51 eV. In the r2SCAN calculations, we obtain $E_{ads}$ of $-$1.61 eV, i.e. equally weak.
Furthermore, we confirmed that there was a tendency for \ce{[Fe^{III}(P)(Cl)]} to move away from the graphene sheet if van der Waals corrections are not included in the calculations.
Without including van der Waals corrections, the resulting distance between \ce{[Fe^{III}(P)(Cl)]} and the graphene sheet increases by 0.81 \AA, compared to the distance (3.26 \AA) measured in the relaxed "bridge 2" structure where the van der Waals correction (DFT-D2) was taken into account.
On the other hand, this distance (3.26 Å) remains nearly unchanged, with an increase of only 0.092 \AA, when the r2SCAN+rVV10 nonlocal vdW-DF functional was employed during structure optimization.
Consequently, the main source of attractive interactions between the porphyrin complex and graphene are van der Waals forces.

\textbf{Electronic structures of the free and physisorbed iron complex.}
The electronic structure of the iron(III) ion in the square-pyramidal ligand field of \ce{[Fe^{III}(P)(Cl)]} is influenced by the axial chloride ligand and the displacement of the iron ion from the porphin ring. The splitting of the $d$ orbital energies is expected to result in a high spin $(xz,yz)^2(xy)^1(z^2)^1(x^2-y^2)^1$ orbital occupation pattern. This highlights the relevance for FeN$_4$-type active sites in single-atom catalysts, where the iron ion is assumed to be situated in a square-planar ligand field of an N-doped graphene plane in its bare state, and have varying axial ligands throughout the catalytic cycle. 

\begin{figure*}
\vskip 2mm
\includegraphics[width=2.0\columnwidth]{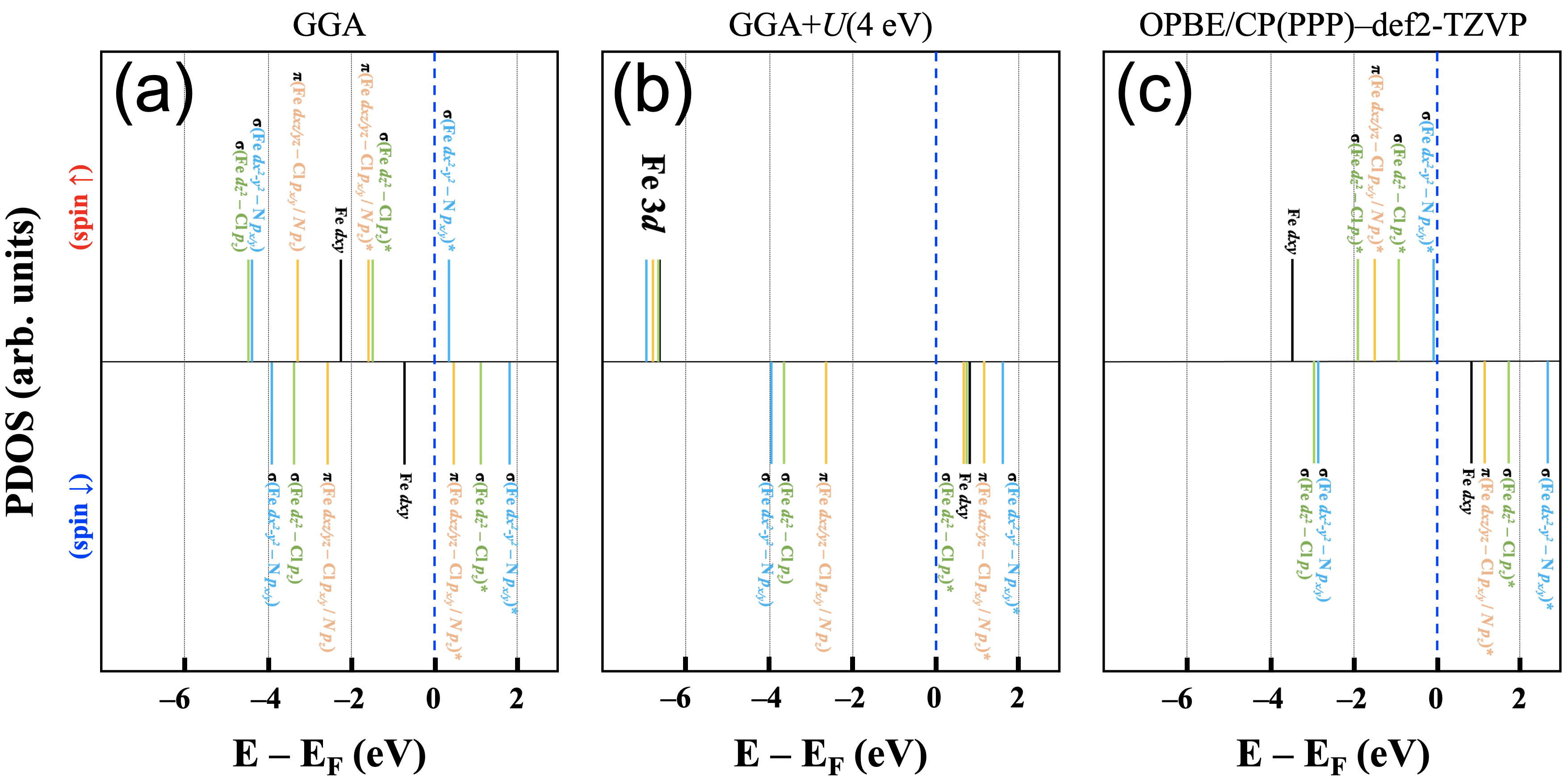}
\caption{
Schematic electronic structures within (a) GGA and (b) GGA + $U$(4 eV) in the \ce{[Fe^{III}(P)(Cl)]} complex adsorbed on a graphene (8$\times$8) sheet. In (c), the electronic structure of \ce{[Fe^{III}(P)(Cl)]} calculated at the OPBE/CP(PPP):def2-TZVP level of theory is shown for comparison.
Each energy state is derived through an analysis of the orbital resolved densities of states.
}
\label{Fe-P_bands}
\end{figure*}

This molecular orbital perspective is reflected in the MO diagram of the complex \ce{[Fe^{III}(P)(Cl)]} calculated with Gaussian basis sets, where the $d$ orbitals span a range of ca. 6 eV (OPBE/CP(PPP):def2-TZVP). The predicted splitting and orbital occupation pattern results purely from the choice of methodology, i.e. chiefly the density functional and basis set.  
On the other hand, in calculations with plane-wave basis sets it is quite helpful to gain insight into understanding the correlation effect arising from the localized $d$ orbitals by varying the Hubbard $U$ term. \cite{Song_2015}
Figure \ref{Fe-P_bands} (a) and (b) show schematic electronic structures predicted with pure GGA and GGA+$U$ for \ce{[Fe^{III}(P)(Cl)]} adsorbed on a graphene (8$\times$8) sheet. The relative energies of each state are based on the results of the orbital projected densities of states.
For comparison, the OPBE/CP(PPP):def2-TZVP electronic structure (Fig. \ref{Fe-P_bands} (c)) is also depicted, wherein all energy states are adjusted relative to the spin down $d_{xy}$ orbital, aligned with the energy level of the corresponding orbital in (b). 

Starting with the pure GGA prediction, we find that aside from the non-bonding Fe $d_{xy}$ orbitals, the Fe 3$d$ orbitals are strongly mixed (i.e., hybridized) with N 2$p$ and Cl 3$p$ character. They hence form bonding and anti-bonding orbitals (i.e., one-electron states) as marked in Figure \ref{Fe-P_bands}.
In the spin-up manifold, all Fe 3$d$ orbitals are occupied except one $\sigma$-type antibonding orbital composed of Fe $d_{x^2-y^2}$ and N $p_{x/y}$ character. In the spin-down manifold, the only occupied orbital with significant iron contributions is the Fe $d_{xy}$ orbital that cannot mix significantly with porphyrin and chloride orbitals. 
As a result, the calculated total spin moment is 3.0 $\mu_{B}$/cell, close to an intermediate spin configuration of $S=3/2$. This spin state is however not the experimentally determined spin state for the system,~\cite{kintner_1991}
therefore appropriate corrections for handling correlation effects in the system need to be considered.

For iron ions, the spatially compact $d$ orbitals indicate significant electron correlation so that a large value of $U$ is needed.~\cite{ferber2012,backes2015,watson2017,Song_2015} We find that values smaller than $U$ = 3 eV result in an intermediate spin state, which is good in agreement with our fixed spin moment results in \ce{[Fe^{III}(P)(Cl)]} as discussed earlier. Using the generally accepted values of $U$ = 4 eV and $J_H$ = 1 eV for iron~\cite{ferber2012,backes2015,watson2017,gga+u_value} produces the desired high spin configuration. As a result, the $d_{xy}$ orbital in the spin-down manifold is unoccupied, whereas all Fe 3d orbitals in the spin-up manifold are fully occupied. For the resulting high spin configuration of $S=5/2$, a calculated total spin moment of 5.0 $\mu_{B}$/cell is found.
Due to the occupied bonding orbitals in the spin-down manifold, ferromagnetic exchange interactions appear between Fe and N(Cl), as shown in the SI, Figure S2. In addition, the calculated spin densities are localized around Fe, N, and Cl, whereas no density appears on the carbon centers.
It is worth noting that iron's high spin configuration can be reached at $U$ larger than 3 eV, whereas $U$ smaller than 3 eV leads to an intermediate spin configuration as observed in the pure GGA results. Similar to the GGA+$U$ results, iron's high spin state is obtained in the r2SCAN calculations, which is however not a routine functional for computational studies in the single atom catalysis field.  
Comparison of this electronic structure (Fig. \ref{Fe-P_bands}(b)) with the molecular orbital diagram (Fig. \ref{Fe-P_bands}(c)) from the calculation with Gaussian-type orbitals shows that the two approaches result in a qualitatively similar picture, though quantitative differences in the orbital ordering and splitting between the spin-up and spin-down manifolds remain.

\section{Discussion and Conclusions} 
In this theoretical study that provides a first step towards bridging the often disparate worlds of molecular chemistry and solid state physics in the \ce{FeN4} catalysis arena, we have shown that the interaction between graphene and an iron(III) complex, \ce{[Fe^{III}(P)(Cl)]}, is negligible. This was demonstrated via the interaction energy, a charge transfer analysis, and analysis of the band structure.
Future studies of Fe-based catalyst models can therefore focus on the single graphene-like layer in which the catalytically active iron ion is embedded.

A central issue for meaningful computational studies of Fe-based catalysts is the correct prediction of the spin states of iron. The spin state of iron is determined by the interplay of the ligand field environment, strong $p-d$ hybridization, and correlation effects. A careful selection of the density functional and additional parameters is therefore needed, which ideally involves comparison with experimental spectroscopy data that are sensitive to details of the electronic structure. \cite{gallenkamp_spectroscopic_2021,ni_2022,gallenkamp_fen_2024,ghosh_spectroscopically_2022,tarrago_experimental_2021}
While spin-orbit coupling will likely be important to obtain a full picture of \ce{FeN4} active sites,\cite{tarrago_experimental_2021} the current theoretical literature in this field rarely even covers the non-relativistic correlation effects adequately. Focusing therefore on parameters that can be included and adjusted more easily for systems similar to or larger than the iron(III) high spin complex \ce{[Fe^{III}(P)(Cl)]}, the ligand field splitting resulting from the electronic structure predictions must be sufficiently low to enable a population of the highest-lying iron d orbital. 

Since the Hubbard $U$ term influences the relative energies of the spin-up and spin-down manifolds, it has a significant influence on the correct prediction of the preferred spin state, as shown here explicitly. With a view beyond the specific system studied here, the predictive power of computational chemistry and physics can only be harnessed if the uncertainty in spin state prediction is known. This is challenging for DFT for most iron complexes, and in addition many examples of molecular complexes with close-lying or even degenerate spin states exist, which determine electronic properties as well as reactivity and catalysis. This work thus raises the question how to choose an appropriate electronic structure description for iron ions at the borderline of molecular and periodic descriptions where the ligand field splitting is expected to be less clear-cut than in the present example, or even completely unknown as it is the case for FeNC catalyst models.

We suggest that molecular and periodic approaches can be used in a complementary manner. Since periodic approaches can be significantly faster than molecular descriptions, once an appropriate Hubbard $U$ value is chosen, they can be used to rapidly evaluate different structures and screen electronic structures, e.g. to evaluate catalytic intermediates. A more detailed electronic structure analysis and the prediction of spectroscopic properties can then be sought with molecular approaches. 

\section*{Acknowledgements}
This work was funded by the Deutsche Forschungsgemeinschaft (DFG, German Research Foundation) -- CRC 1487, ``Iron, upgraded!'' -- project number 443703006. 
Computations for this work were conducted on the Lichtenberg II high performance computer at TU Darmstadt.

\section*{Author Contributions}
Y-J S., C.G. and G. Ll. performed the calculations. V. K. and R. V.  conceived and supervised the project. All authors contributed to the discussions and
the writing of the paper.

\section*{Conflicts of interest}
There are no conflicts to declare.

\bibliography{main} 

\end{document}

\title{Supporting Information \\ for \\ Influence of graphene on the electronic and magnetic properties of an iron(III) porphyrin chloride complex}

\author{Young-Joon Song}
\affiliation{Institut f\"ur Theoretische Physik, Goethe-Universit\"at Frankfurt, Max-von-Laue-Str. 1, 60438 Frankfurt am Main, Germany}

\author{Charlotte Gallenkamp}
\affiliation{TU Darmstadt, Department of Chemistry, Quantum Chemistry, Peter-Gr\"unberg-Str. 4, 64287 Darmstadt, Germany}

\author{Gen\'{i}s Lleopart} 
\affiliation{Departament de Ci\'encia de Materials i Qu\'imica F\'isica and Institut de Qu\'imica Te\'orica i Computacional (IQTC), Universitat de Barcelona, c/ Mart\'i i Franqu\'es 1-11, 08028 Barcelona, Spain}

\author{Vera Krewald}
\affiliation{TU Darmstadt, Department of Chemistry, Quantum Chemistry, Peter-Gr\"unberg-Str. 4, 64287 Darmstadt, Germany}

\author{Roser Valent\'{i}}
\affiliation{Institut f\"ur Theoretische Physik, Goethe-Universit\"at Frankfurt, Max-von-Laue-Str. 1, 60438 Frankfurt am Main, Germany}

\maketitle

\begin{figure}[tb]
\vskip 2mm
\includegraphics[width=0.99\columnwidth]{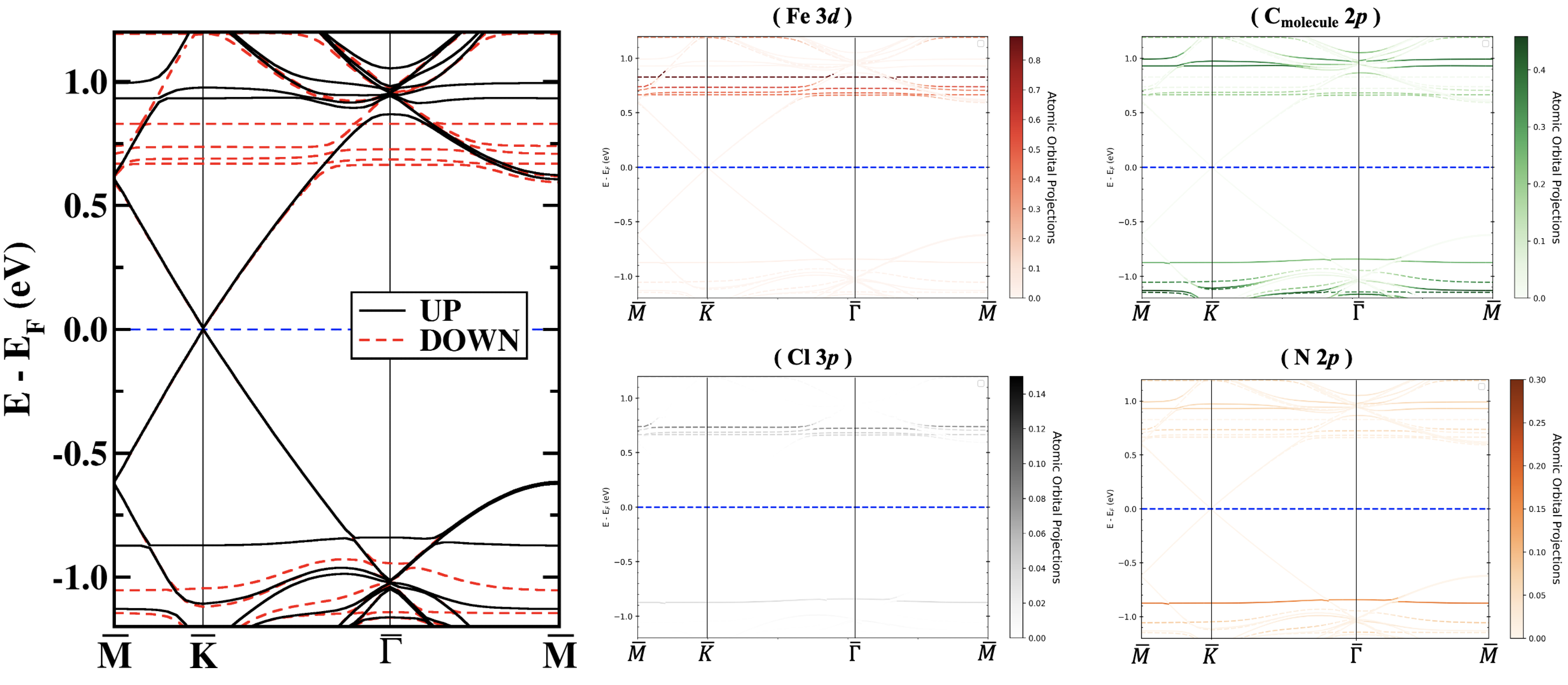}
\caption{
(Left) Spin-resolved and (middle and right) both spin- and orbital-resolved band structures in \ce{[Fe^{III}(P)(Cl)]} absorbed on a graphene sheet (8$\times$8) within GGA + $U$ (4 eV).
In all plots, solid (dashed) lines represent spin-up (spin-down) bands.
The main contribution to forming the Dirac point at the Fermi level stems from C$_{graphene}$ 2$p$, which is shown in the main text.
}
\label{fatbands}
\end{figure}

\begin{figure}[b]
\vskip 2mm
\includegraphics[width=0.99\columnwidth]{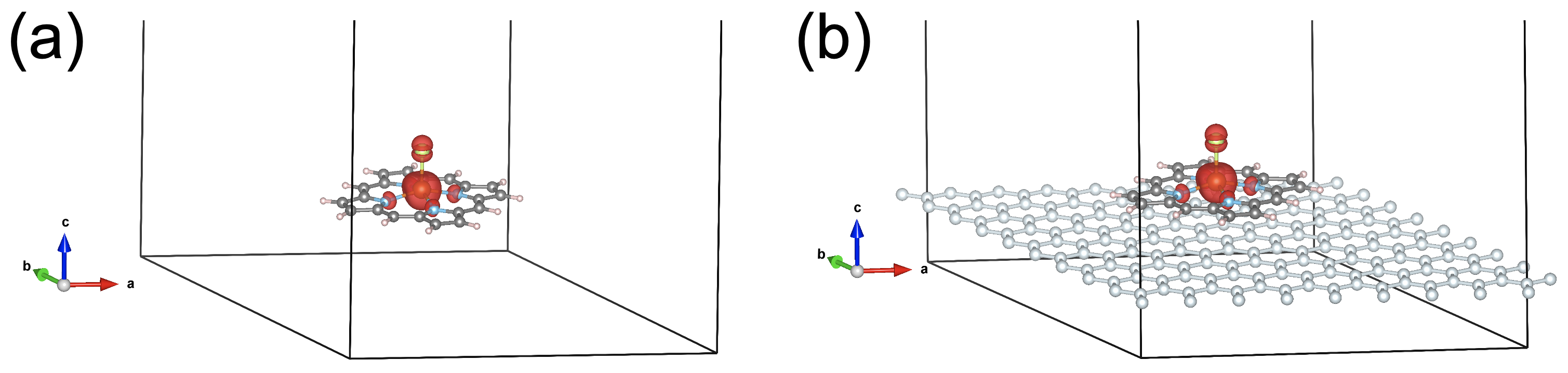}
\caption{
Spin density plot ($\rho_\uparrow$ $-$ $\rho_\downarrow$) of (a) \ce{[Fe^{III}(P)(Cl)]} and (b) \ce{[Fe^{III}(P)(Cl)]} absorbed on a graphene sheet (8$\times$8) within GGA + $U$ (4 eV) with an isovalue of 0.067 $e$/\AA. The spin density plots of both cases are almost similar independent of the existence of graphene, indicating a minimal role of graphene.
}
\label{Fe-P_sd}
\end{figure}

\newpage

\begin{figure*}[t]
\vskip 2mm
\includegraphics[width=0.99\columnwidth]{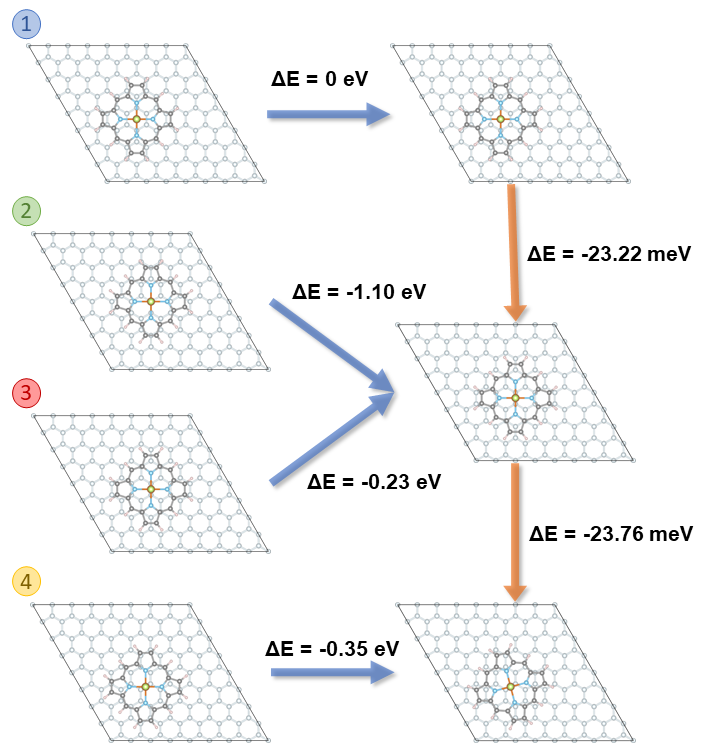}
\caption{
Structures before (in the left column) and after (in the right column) relaxation with VASP. Blue arrows highlight variations in total energy between the unrelaxed and relaxed structures, while orange arrows denote differences in total energy between two relaxed structures. As a result, the case04 (named bridge2 in the main text) is found to be energetically the most stable.
}
\label{Fe-P_rel}
\end{figure*}

\newpage

\begin{table}
\centering
 \begin{tabular}{|c|c|c|c|c|c|c|c|c|} \hline
  Atom  & $x$ & $y$ & $z$ & \hspace{20pt} & Atom  & $x$ & $y$ & $z$ \\ \hline
      Fe & -0.008727 & -0.018712 & -0.027386 & & H & -2.807025 & 4.920590 & 0.764329 \\\hline
      N & 2.039261 & -0.038703 & -0.423795 & & H & -2.918084 & 6.613482 & 0.199272 \\\hline
      N & 0.018645 & -2.085525 & -0.291067 & & H & -1.434104 & 6.035505 & 1.013858 \\\hline
      N & -2.027350 & -0.079449 & -0.558443 & & H & 1.142692 & 6.135294 & 1.054967 \\\hline
      N & -0.006855 & 1.965865 & -0.690620 & & H & 2.695939 & 6.711881 & 0.380044 \\\hline
      H & 5.258322 & 2.520041 & -0.103633 & & H & 2.548809 & 5.036769 & 0.988904 \\\hline
      H & 6.329694 & 1.112051 & -0.219948 & & C & 2.866594 & 1.066465 & -0.540529 \\\hline
      H & 5.285019 & -2.436879 & 0.425613 & & C & 4.261354 & 0.663259 & -0.499534 \\\hline
      H & 6.341749 & -1.072927 & 0.019912 & & C & 5.436649 & 1.584783 & -0.664586 \\\hline
      H & 3.230297 & -3.218110 & -0.085206 & & C & 4.269658 & -0.714215 & -0.350769 \\\hline
      H & 1.136046 & -6.341818 & 0.354381 & & C & 5.454771 & -1.637094 & -0.317767 \\\hline
      H & 2.503704 & -5.247820 & 0.628502 & & C & 2.880382 & -1.133481 & -0.307521 \\\hline
      H & -2.467761 & -5.296082 & 0.427187 & & C & 2.452553 & -2.456629 & -0.181559 \\\hline
      H & -1.055603 & -6.359770 & 0.295138 & & C & 1.130690 & -2.905653 & -0.192492 \\\hline
      H & -3.176103 & -3.282842 & -0.304438 & & C & 0.729065 & -4.298698 & -0.112836 \\\hline
      H & -6.303957 & -1.247590 & -0.918514 & & C & 1.669406 & -5.469322 & -0.061595 \\\hline
      H & -5.222178 & -2.650110 & -1.001224 & & C & -0.655328 & -4.313684 & -0.164380 \\\hline
      H & -6.317108 & 0.939204 & -1.144815 & & C & -1.569199 & -5.506158 & -0.180724 \\\hline
      H & -5.252174 & 2.305824 & -1.519959 & & C & -1.080201 & -2.929031 & -0.272717 \\\hline
      H & -3.216763 & 3.093999 & -0.948123 & & C & -2.408895 & -2.505552 & -0.344843 \\\hline
      H & -2.482437 & 5.098214 & -1.742792 & & C & -2.854273 & -1.188370 & -0.476548 \\\hline
      H & -1.115283 & 6.199207 & -1.493041 & & C & -4.248183 & -0.790110 & -0.564417 \\\hline
      H & 1.075649 & 6.207170 & -1.470083 & & C & -5.424473 & -1.717230 & -0.444499 \\\hline
      H & 2.485743 & 5.135638 & -1.547798 & & C & -4.256295 & 0.587876 & -0.708390 \\\hline
      H & 3.189948 & 3.157312 & -0.747055 & & C & -5.442666 & 1.506624 & -0.781002 \\\hline
      H & 5.944713 & 1.012574 & -2.718194 & & C & -2.867703 & 1.012174 & -0.702502 \\\hline
      H & 6.594420 & 2.602857 & -2.218403 & & C & -2.439769 & 2.335124 & -0.827807 \\\hline
      H & 4.858113 & 2.425058 & -2.607243 & & C & -1.117612 & 2.783421 & -0.820540 \\\hline
      H & 4.889825 & -2.867887 & -2.044871 & & C & -0.714390 & 4.171741 & -0.957215 \\\hline
      H & 6.624945 & -2.949758 & -1.620923 & & C & -1.652964 & 5.340886 & -1.053909 \\\hline
      H & 5.970159 & -1.503011 & -2.444220 & & C & 0.670115 & 4.186312 & -0.912289 \\\hline
      H & 1.437054 & -6.103705 & -2.147584 & & C & 1.586289 & 5.376204 & -0.952695 \\\hline
      H & 2.930247 & -6.697610 & -1.362115 & & C & 1.092877 & 2.806321 & -0.748791 \\\hline
      H & 2.806520 & -4.989608 & -1.876075 & & C & 2.421774 & 2.383066 & -0.677298 \\\hline
      H & -2.537881 & -5.076433 & -2.101214 & & C & 5.725186 & 1.926615 & -2.139326 \\\hline
      H & -2.678426 & -6.779208 & -1.573367 & & C & 5.751916 & -2.277069 & -1.687923 \\\hline
      H & -1.128494 & -6.166191 & -2.223632 & & C & 2.244333 & -5.836192 & -1.443668 \\\hline
      H & -4.906452 & -2.556939 & 1.514843 & & C & -2.004195 & -5.905364 & -1.604038 \\\hline
      H & -6.627184 & -2.747762 & 1.066698 & & C & -5.761194 & -2.064504 & 1.018659 \\\hline
      H & -6.007300 & -1.153390 & 1.591422 & & C & -5.783396 & 2.147779 & 0.578480 \\\hline
      H & -6.018392 & 1.373724 & 1.329746 & & C & -2.237331 & 5.751486 & 0.311763 \\\hline
      H & -6.658364 & 2.814723 & 0.484857 & & C & 2.019075 & 5.842390 & 0.450813 \\\hline
      H & -4.935479 & 2.744545 & 0.958776 & & Cl & -0.097459 & 0.204579 & 2.223812 \\\hline
\end{tabular}
    \caption{Cartesian coordinates (\AA) of the relaxed \ce{[Fe^{III}(OEP)(Cl)]} complex in the high spin state at the TPSS/def2-TZVP:def2-SVP level of theory.}
    \label{orca_coord_Fe-P}
\end{table}

\begin{table}
\centering
 \begin{tabular}{|c|c|c|c|c|c|c|c|c|} \hline
  Atom  & $x$ & $y$ & $z$ & \hspace{20pt} & Atom  & $x$ & $y$ & $z$ \\ \hline
      Fe  &  -0.007323  &  -0.018718  &  -0.007826  & &  C  &  4.257218  &  -0.706805  &  -0.352841  \\\hline
      N  &  2.038146  &  -0.039090  &  -0.429355  & &  C  &  2.878673  &  -1.134982  &  -0.317916  \\\hline
      N  &  0.018471  &  -2.084489  &  -0.294025  & &  C  &  2.456509  &  -2.461961  &  -0.202286  \\\hline
      N  &  -2.028194  &  -0.076178  &  -0.539960  & &  C  &  1.131260  &  -2.904443  &  -0.198766  \\\hline
      N  &  -0.008871  &  1.967210  &  -0.673518  & &  C  &  0.720997  &  -4.285812  &  -0.107203  \\\hline
      H  &  3.235136  &  -3.225119  &  -0.120551  & &  C  &  -0.651769  &  -4.298816  &  -0.147407  \\\hline
      H  &  -3.180521  &  -3.285247  &  -0.304296  & &  C  &  -1.082051  &  -2.925356  &  -0.262768  \\\hline
      H  &  -3.222962  &  3.102994  &  -0.915146  & &  C  &  -2.412960  &  -2.507474  &  -0.340975  \\\hline
      H  &  3.191959  &  3.161897  &  -0.745521  & &  C  &  -2.852643  &  -1.188181  &  -0.477382  \\\hline
      H  &  -1.325078  &  -5.156490  &  -0.106919  & &  C  &  -4.235110  &  -0.785976  &  -0.587107  \\\hline
      H  &  1.407038  &  -5.130532  &  -0.027158  & &  C  &  -4.244523  &  0.581107  &  -0.717769  \\\hline
      H  &  5.115075  &  -1.377890  &  -0.287227  & &  C  &  -2.867781  &  1.015678  &  -0.687330  \\\hline
      H  &  5.096650  &  1.342438  &  -0.555429  & &  C  &  -2.445652  &  2.342500  &  -0.802343  \\\hline
      H  &  1.339258  &  5.024517  &  -1.015375  & &  C  &  -1.120112  &  2.784029  &  -0.804729  \\\hline
      H  &  -1.393256  &  4.999250  &  -1.087535  & &  C  &  -0.708365  &  4.159332  &  -0.960824  \\\hline
      H  &  -5.101101  &  1.247547  &  -0.830421  & &  C  &  0.664333  &  4.171983  &  -0.924518  \\\hline
      H  &  -5.082199  &  -1.473586  &  -0.570505  & &  C  &  1.092490  &  2.804399  &  -0.746446  \\\hline
      C  &  2.863811  &  1.068460  &  -0.533741  & &  C  &  2.423922  &  2.387111  &  -0.673466  \\\hline
      C  &  4.248017  &  0.659896  &  -0.487807  & &  Cl  &  -0.071830  &  0.185690  &  2.229841  \\\hline
\end{tabular}
    \caption{Cartesian coordinates (\AA) of the relaxed \ce{[Fe^{III}(P)(Cl)]} complex in the high spin state at the TPSS/def2-TZVP:def2-SVP level of theory.}
    \label{orca_coord_Fe-P_trunc}
\end{table}

\begin{table}
    \centering
    \begin{tabular}{|c|c|c|c|c|c|c|c|c|c|c|c|c|c|} \hline 
            Atom  & $x$ & $y$ & $z$ & \hspace{10pt} & Atom  & $x$ & $y$ & $z$ & \hspace{10pt} & Atom  & $x$ & $y$ & $z$\\ \hline
      Fe & 0.416965 & 0.396253 & 0.274903 & & C & 0.250000 & 0.250000 & 0.130769 & & C & 0.208333 & 0.166667 & 0.130769 \\\hline
      N & 0.531945 & 0.426209 & 0.256818 & & C & 0.250000 & 0.375000 & 0.130769 & & C & 0.208333 & 0.291667 & 0.130769 \\\hline
      N & 0.448830 & 0.511718 & 0.257262 & & C & 0.250000 & 0.500000 & 0.130769 & & C & 0.208333 & 0.416667 & 0.130769 \\\hline
      N & 0.301986 & 0.366302 & 0.256820 & & C & 0.250000 & 0.625000 & 0.130769 & & C & 0.208333 & 0.541667 & 0.130769 \\\hline
      N & 0.385105 & 0.280794 & 0.257259 & & C & 0.250000 & 0.750000 & 0.130769 & & C & 0.208333 & 0.666667 & 0.130769 \\\hline
      H & 0.107898 & 0.244178 & 0.253478 & & C & 0.250000 & 0.875000 & 0.130769 & & C & 0.208333 & 0.791667 & 0.130769 \\\hline
      H & 0.150567 & 0.398669 & 0.252731 & & C & 0.375000 & 0.000000 & 0.130769 & & C & 0.208333 & 0.916667 & 0.130769 \\\hline
      H & 0.419822 & 0.665306 & 0.254978 & & C & 0.375000 & 0.125000 & 0.130769 & & C & 0.333333 & 0.041667 & 0.130769 \\\hline
      H & 0.573608 & 0.705341 & 0.255507 & & C & 0.375000 & 0.250000 & 0.130769 & & C & 0.333333 & 0.166667 & 0.130769 \\\hline
      H & 0.726036 & 0.548330 & 0.253480 & & C & 0.375000 & 0.375000 & 0.130769 & & C & 0.333333 & 0.291667 & 0.130769 \\\hline
      H & 0.683362 & 0.393836 & 0.252750 & & C & 0.375000 & 0.500000 & 0.130769 & & C & 0.333333 & 0.416667 & 0.130769 \\\hline
      H & 0.260330 & 0.087172 & 0.255475 & & C & 0.375000 & 0.625000 & 0.130769 & & C & 0.333333 & 0.541667 & 0.130769 \\\hline
      H & 0.414116 & 0.127208 & 0.254958 & & C & 0.375000 & 0.750000 & 0.130769 & & C & 0.333333 & 0.666667 & 0.130769 \\\hline
      H & 0.286281 & 0.530836 & 0.255147 & & C & 0.375000 & 0.875000 & 0.130769 & & C & 0.333333 & 0.791667 & 0.130769 \\\hline
      H & 0.647916 & 0.624946 & 0.256345 & & C & 0.500000 & 0.000000 & 0.130769 & & C & 0.333333 & 0.916667 & 0.130769 \\\hline
      H & 0.547653 & 0.261675 & 0.255158 & & C & 0.500000 & 0.125000 & 0.130769 & & C & 0.458333 & 0.041667 & 0.130769 \\\hline
      H & 0.186018 & 0.167565 & 0.256340 & & C & 0.500000 & 0.250000 & 0.130769 & & C & 0.458333 & 0.166667 & 0.130769 \\\hline
      Cl & 0.416964 & 0.396250 & 0.359825 & & C & 0.500000 & 0.375000 & 0.130769 & & C & 0.458333 & 0.291667 & 0.130769 \\\hline
      C & 0.317779 & 0.498368 & 0.256070 & & C & 0.500000 & 0.500000 & 0.130769 & & C & 0.458333 & 0.416667 & 0.130769 \\\hline
      C & 0.592239 & 0.569812 & 0.256915 & & C & 0.500000 & 0.625000 & 0.130769 & & C & 0.458333 & 0.541667 & 0.130769 \\\hline
      C & 0.516156 & 0.294143 & 0.256076 & & C & 0.500000 & 0.750000 & 0.130769 & & C & 0.458333 & 0.666667 & 0.130769 \\\hline
      C & 0.241696 & 0.222698 & 0.256912 & & C & 0.500000 & 0.875000 & 0.130769 & & C & 0.458333 & 0.791667 & 0.130769 \\\hline
      C & 0.238015 & 0.291593 & 0.256418 & & C & 0.625000 & 0.000000 & 0.130769 & & C & 0.458333 & 0.916667 & 0.130769 \\\hline
      C & 0.272581 & 0.416786 & 0.255880 & & C & 0.625000 & 0.125000 & 0.130769 & & C & 0.583333 & 0.041667 & 0.130769 \\\hline
      C & 0.399475 & 0.542419 & 0.256739 & & C & 0.625000 & 0.250000 & 0.130769 & & C & 0.583333 & 0.166667 & 0.130769 \\\hline
      C & 0.524109 & 0.574877 & 0.257143 & & C & 0.625000 & 0.375000 & 0.130769 & & C & 0.583333 & 0.291667 & 0.130769 \\\hline
      C & 0.309825 & 0.217634 & 0.257134 & & C & 0.625000 & 0.500000 & 0.130769 & & C & 0.583333 & 0.416667 & 0.130769 \\\hline
      C & 0.434459 & 0.250094 & 0.256736 & & C & 0.625000 & 0.625000 & 0.130769 & & C & 0.583333 & 0.541667 & 0.130769 \\\hline
      C & 0.561350 & 0.375724 & 0.255886 & & C & 0.625000 & 0.750000 & 0.130769 & & C & 0.583333 & 0.666667 & 0.130769 \\\hline
      C & 0.595918 & 0.500918 & 0.256416 & & C & 0.625000 & 0.875000 & 0.130769 & & C & 0.583333 & 0.791667 & 0.130769 \\\hline
      C & 0.166577 & 0.295085 & 0.254748 & & C & 0.750000 & 0.000000 & 0.130769 & & C & 0.583333 & 0.916667 & 0.130769 \\\hline
      C & 0.188033 & 0.372759 & 0.254381 & & C & 0.750000 & 0.125000 & 0.130769 & & C & 0.708333 & 0.041667 & 0.130769 \\\hline
      C & 0.444720 & 0.626925 & 0.256131 & & C & 0.750000 & 0.250000 & 0.130769 & & C & 0.708333 & 0.166667 & 0.130769 \\\hline
      C & 0.522060 & 0.647056 & 0.256446 & & C & 0.750000 & 0.375000 & 0.130769 & & C & 0.708333 & 0.291667 & 0.130769 \\\hline
      C & 0.667355 & 0.497424 & 0.254750 & & C & 0.750000 & 0.500000 & 0.130769 & & C & 0.708333 & 0.416667 & 0.130769 \\\hline
      C & 0.645899 & 0.419750 & 0.254391 & & C & 0.750000 & 0.625000 & 0.130769 & & C & 0.708333 & 0.541667 & 0.130769 \\\hline
      C & 0.311877 & 0.145457 & 0.256425 & & C & 0.750000 & 0.750000 & 0.130769 & & C & 0.708333 & 0.666667 & 0.130769 \\\hline
      C & 0.389217 & 0.165588 & 0.256117 & & C & 0.750000 & 0.875000 & 0.130769 & & C & 0.708333 & 0.791667 & 0.130769 \\\hline
      C & 0.000000 & 0.000000 & 0.130769 & & C & 0.875000 & 0.000000 & 0.130769 & & C & 0.708333 & 0.916667 & 0.130769 \\\hline
      C & 0.000000 & 0.125000 & 0.130769 & & C & 0.875000 & 0.125000 & 0.130769 & & C & 0.833333 & 0.041667 & 0.130769 \\\hline
      C & 0.000000 & 0.250000 & 0.130769 & & C & 0.875000 & 0.250000 & 0.130769 & & C & 0.833333 & 0.166667 & 0.130769 \\\hline
      C & 0.000000 & 0.375000 & 0.130769 & & C & 0.875000 & 0.375000 & 0.130769 & & C & 0.833333 & 0.291667 & 0.130769 \\\hline
      C & 0.000000 & 0.500000 & 0.130769 & & C & 0.875000 & 0.500000 & 0.130769 & & C & 0.833333 & 0.416667 & 0.130769 \\\hline
      C & 0.000000 & 0.625000 & 0.130769 & & C & 0.875000 & 0.625000 & 0.130769 & & C & 0.833333 & 0.541667 & 0.130769 \\\hline
      C & 0.000000 & 0.750000 & 0.130769 & & C & 0.875000 & 0.750000 & 0.130769 & & C & 0.833333 & 0.666667 & 0.130769 \\\hline
      C & 0.000000 & 0.875000 & 0.130769 & & C & 0.875000 & 0.875000 & 0.130769 & & C & 0.833333 & 0.791667 & 0.130769 \\\hline
      C & 0.125000 & 0.000000 & 0.130769 & & C & 0.083333 & 0.041667 & 0.130769 & & C & 0.833333 & 0.916667 & 0.130769 \\\hline
      C & 0.125000 & 0.125000 & 0.130769 & & C & 0.083333 & 0.166667 & 0.130769 & & C & 0.958333 & 0.041667 & 0.130769 \\\hline
      C & 0.125000 & 0.250000 & 0.130769 & & C & 0.083333 & 0.291667 & 0.130769 & & C & 0.958333 & 0.166667 & 0.130769 \\\hline
      C & 0.125000 & 0.375000 & 0.130769 & & C & 0.083333 & 0.416667 & 0.130769 & & C & 0.958333 & 0.291667 & 0.130769 \\\hline
      C & 0.125000 & 0.500000 & 0.130769 & & C & 0.083333 & 0.541667 & 0.130769 & & C & 0.958333 & 0.416667 & 0.130769 \\\hline
      C & 0.125000 & 0.625000 & 0.130769 & & C & 0.083333 & 0.666667 & 0.130769 & & C & 0.958333 & 0.541667 & 0.130769 \\\hline
      C & 0.125000 & 0.750000 & 0.130769 & & C & 0.083333 & 0.791667 & 0.130769 & & C & 0.958333 & 0.666667 & 0.130769 \\\hline
      C & 0.125000 & 0.875000 & 0.130769 & & C & 0.083333 & 0.916667 & 0.130769 & & C & 0.958333 & 0.791667 & 0.130769 \\\hline
      C & 0.250000 & 0.000000 & 0.130769 & & C & 0.208333 & 0.041667 & 0.130769 & & C & 0.958333 & 0.916667 & 0.130769 \\\hline
      C & 0.250000 & 0.125000 & 0.130769 & & & & & & & & & & \\\hline
          \end{tabular}
    \caption{Fractional coordinates of the relaxed \ce{[Fe^{III}(P)(Cl)]} complex absorbed graphene (named bridge 2 in the main text) within GGA + $U$ ($U$ = 4 eV) with the lattice parameters of $a_H$ = 19.68 \AA~and $c_H$ = 26 \AA~in VASP. 
    Here $H$ denotes the hexagonal unit cell with $\overrightarrow{a_1}$=$a_H$(1,0,0), 
    $\overrightarrow{a_2}$=$a_H$($-\frac{1}{2}$,$\frac{\sqrt{3}}{2}$,0), 
    and $\overrightarrow{a_3}$=$c_H$(0,0,1).}
    \label{str_coord_Fe-P}
\end{table}

\begin{table}
    \centering
    \begin{tabular}{|c|c|c|c|c|c|c|}   
     \hline
     %\vspace{-1mm} \\ 
     \multicolumn{7}{|c|}{INCAR tags} \\
     %\vspace{-1mm} \\
     \hline\hline
     ISTART= 0 & \hspace{10pt} & EDIFF= 1.0E-6 & \hspace{10pt} & BMIX\_MAG= 0.00001 & \hspace{10pt} & SIGMA= 0.05 \\\hline
     ICHARG= 2 & & ISPIN= 2 & & LDAU= .TRUE.    & & IVDW = 1 \\\hline
     INIWAV= 1 & & MAGMOM=  5.0 165*0.0 & & LDAUTYPE= 1    & & BMIX= 0.00001 \\\hline
     ENCUT= 500.00 eV & & LORBIT= 11 & & LDAUL= 2 10*-1 & & AMIX\_MAG= 0.8\\\hline
     PREC= Accurate & & ISYM= 2 & & LDAUU= 4.0 10*0.0   & & LDAUPRINT = 2  \\\hline
     ISMEAR= 0 & & AMIX= 0.2 & & LDAUJ= 1.0 10*0.0   & & LMAXMIX= 4 \\\hline
    \end{tabular}
    \caption{Details of INCAR tags used in the VASP calculations. INCAR is one of the vital input files in VASP. }
    \label{incar}
\end{table}